\documentclass[aps,prl,reprint,twocolumn,superscriptaddress]{revtex4-2}
\usepackage{amsmath}
\usepackage{amssymb}
\usepackage[paper=a4paper,left=25mm,right=25mm,top=25mm,bottom=25mm]{geometry}
\usepackage{mathtools}
\usepackage{dsfont}
\usepackage{hyperref} 
\usepackage{cleveref}
\crefname{figure}{Fig.}{Fig.}
\crefname{equation}{Eq.}{Eq.}
\usepackage{siunitx}
\usepackage{subcaption}
\usepackage{graphicx} 
\usepackage[sort&compress]{natbib}
\graphicspath{{Graphen}}
\usepackage{float}
\begin{document}

\title{Exact Results for the Ericson Transition in Stochastic Quantum Scattering and Experimental Validation}
\author{Simon Köhnes}\email{simon.koehnes@uni-due.de}\affiliation{Fakult\"at f\"ur Physik, Universit\"at Duisburg-Essen, Duisburg, Germany}
\author{Jiongning Che}
\altaffiliation[now at: ]{Yangtze Delta Region Institute, University of Electronic Science and Technology of China, Huzhou, China}
\affiliation{Lanzhou Center for Theoretical Physics, Lanzhou University, Lanzhou, Gansu, China}
\author{Barbara Dietz} \email{bdietzp@gmail.com}
\affiliation{Lanzhou Center for Theoretical Physics, Lanzhou University, Lanzhou, Gansu, China}
\affiliation{Institute for Basic Science and Korea University of Science and Technology, Daejeon, Korea}
\author{Thomas Guhr} \email{thomas.guhr@uni-due.de} 
\affiliation{Fakult\"at f\"ur Physik, Universit\"at Duisburg-Essen, Duisburg, Germany}
\date{\today}

\begin{abstract}
At lower energies, the resonances in scattering experiments are often isolated. The crucial parameter is the ratio of average resonance width and average mean level spacing. Towards larger energies, this parameter grows, because the resonances overlap. Eventually the cross--section becomes a random function and the scattering matrix elements follow a universal Gaussian distribution. For more than sixty years, this Ericson transition awaits a concise analytical treatment. We provide a complete solution within the Heidelberg approach which provides a full-fledged model of the scattering process. As a side result, we obtain explicit formulae
for the moments of the distributions. We compare with microwave experiments.
\end{abstract}
\maketitle
\indent\textit{Introduction} ---
 Scattering experiments are ubiquitous in quantum and classical physics. Often, the measurements show, in a broad sense, chaotic or, more generally, stochastic dynamics. Numerous examples are found in nuclear  \cite{BRINK196377, VONBRENTANO1964, Porter1965,HARNEY199035,Main1992,  Feshbach1993, Concepts,Weidenmüller, Weidenmüller2,Frisch2014, Kawano2015}, atomic and molecular physics \cite{LOMBARDI1993, GREMAUD1993, DUPRET1995, SCHINKE1995, Scott1996, PhysRevLett.95.194101,PhysRevLett.95.263601,Mayle2013}, quantum graphs \cite{Quantengraphen, QuantumChaos, QuantumChaos2} microwave billiards \cite{ PhysRevLett.65.3072,PhysRevLett.74.62,PhysRevLett.94.144101, PhysRevE.74.036213, PhysRevE.81.036205,  Dietz2010}, microwave networks and cavities \cite{PhysRevLett.100.254101,Hul2005,Lawniczak2008,Lawniczak2020,Chen2021} and classical wave systems \cite{Weaver,PhysRevLett.75.1546, GROS2014664}. Theory of electronic transport \cite{RevModPhys.69.731,PhysRevB.35.1039,RevModPhys.72.895, PhysRevLett.69.1584, MIRLIN1994325, Folk1996,BringiChandrasekar2001,  Yeh2012} is closely related. 
The field of stochastic quantum scattering originated in nuclear physics and was pioneered by Ericson in the 1960s \cite{Ericson1, Ericson2, Ericson3}. He showed that the cross--section versus energy becomes a random function when the resonances strongly overlap. The Ericson regime was first identified in nuclei  \cite{Ericson1, Ericson2, Ericson3}, see a review in \cite{Weidenmüller2}, and subsequently in numerous systems, \textit{e.g.}, in electronic transport \cite{HAW90,SICZB09}, in atoms \cite{Bluemel96,GD97,PhysRevLett.95.194101,PhysRevLett.95.263601,EM09,EM10,EKM11}, in ultracold
atomic-molecular systems~\cite{MRB12}, in Bose--Einstein condensates \cite{HHCKB12} and quantum graphs \cite{ZM13}.

\textit{Scattering process} ---
The scattering process is encoded in the complex and unitary scattering matrix $S$ \cite{Mahaux, AGASSI1975,VWZ}. If $M$ channels are connected to the interaction zone described by a Hermitean Hamilton operator, the elements $S_{ab}$ of the scattering matrix are
\begin{align}
\nonumber S_{ab} &= \delta_{ab} - 2\pi i W_a^\dagger G W_b.\\\label{eq:Heidelberg}
\end{align}
In a proper basis of Hilbert space one may use a matrix notation such that
\begin{align}
G &=\left( E\mathds{1}_N - H +i\pi\sum\limits_{c=1}^{M} W_c W_c^\dagger\right)^{-1}
\end{align}
is the $N\times N$ matrix resolvent at energy $E$. Eventually the limit $N\to \infty$ has to be taken. Here, $H$ is the Hamilton matrix and the $N$-component vectors $W_c,\ c=1,...,M$ describe the coupling between bound states in the interaction zone and the channels. One may assume their orthogonality.  Apart from kinematic factors, the cross--sections are given by $\sigma_{ab} = |S_{ab}|^2$. \\
\indent\textit{Heidelberg approach} ---
In a generic chaotic, complex or stochastic setting, $H$ can be chosen as a random matrix, drawn from one of the Gaussian Ensembles GOE $(\beta=1)$, GUE $(\beta=2)$ or GSE $(\beta = 4)$, where $\beta= 1,4$ describe time-reversal invariant systems. Here, we focus on the time-reversal non-invariant unitary $(\beta=2)$ case of Hermitean $N\times N$ matrices $H$ drawn from a Gaussian probability density with variance $\nu^2/N$. The cases $\beta=1,4$ will be treated elsewhere. \\
\indent\textit{Ericson regime and Weisskopf estimate} ---
The Ericson transition is characterized by the ratio of average resonance width $\Gamma$ and average mean level spacing $D$, i.e. by the dimensionless parameter
\begin{align}
\Xi=\frac{\Gamma}{D}= \frac{1}{2\pi} \sum\limits_{c=1}^{M}T_c, 
\end{align}
where the second equation is known as the Weisskopf estimate \cite{Weisskopf} and $T_c$ are the transmission coefficients. We have two goals: First, we want to prove that the real and imaginary parts of $S_{ab}$ for $a \neq b$ are Gaussian distributed in the Ericson regime. The previous reasoning was mainly heuristic \cite{AGASSI1975}. Second, we want to provide asymptotic corrections to the Gaussian form, thereby fully clarifying the transition to the Ericson regime.

 \textit{Supersymmetry and starting point} ---
Some years ago, a variant of the supersymmetry method \cite{Efetov,VWZ,Efetov1996} was put forward in a series of papers \cite{PhysRevLett.111.030403, Matrixelemente, Wirkungsquerschnitte}, which made it possible to explicitly calculate the distributions of $\text{Re}(S_{ab}),\  \text{Im}(S_{ab})$ and $\sigma_{ab}$ in terms of low-dimensional integrals. In the sequel, we employ those results to study the Ericson transition by means of asymptotic expansions in powers of $1/ \Xi$. Previously, only the correlator of the $S$-matrix for $\beta=1$ could be investigated in this way \cite{VerbaarschotCorr}. \\ 
\indent\textit{Derivation of the Gaussian} --- We use the characteristic function to generate the moments which we then expand asymptotically in $1/\Xi$. The characteristic function in the unitary case is given by \cite{PhysRevLett.111.030403,Matrixelemente}
\begin{align}\nonumber
R_s(k) =\ &1 -\int\limits_{1}^{\infty} d\lambda_1\int\limits_{-1}^{1} d\lambda_2 \frac{k^2}{4 (\lambda_1 -\lambda_2)^2}\mathcal{F}_U(\lambda_1,\lambda_2)\\\label{eq:Sub1}
    &\times \left( t_a^1t_b^1+t_a^2t_b^2\right) J_0\left( k \sqrt{ t_a^1t_b^1}\right),
\end{align}where $s=1,2$ stands for real and imaginary part, $x_1= \text{Re}(S_{ab}),  x_2 = \text{Im}(S_{ab})$, $t_c^j = \sqrt{|\lambda_j^2-1|}/(g_c^++\lambda_j), g_c^+ = 2/T_c-1$ and $J_0(k)$ denotes the Bessel function of order zero.  We notice that the distributions and thus the characteristic functions for $s=1,2$ are equal \cite{PhysRevLett.111.030403, Matrixelemente}. We still keep the index $s$. The channel factor is given by 
\begin{align}
\mathcal{F}_U(\lambda_1,\lambda_2) = \prod\limits_{c=1}^{M} \frac{g_c^+ + \lambda_2}{g_c^++\lambda_1}.
\end{align}
As seen in \cref{eq:Sub1}, the integrand has a coordinate singularity at $\lambda_1=\lambda_2=1$. This term is rooted in a singular value decomposition of a supermatrix and the resulting integration measure. Such singularities are frequently encountered when applying the supersymmetry method
 \cite{VWZ, Efetov,VerbaarschotCorr, PLUHAR19951,Efetov1996,Fyodorov, Rozhkov2003,Rozhkov2004,PhysRevLett.111.030403, Matrixelemente, Wirkungsquerschnitte}. To remove this singularity, we change variables,
$\lambda_1 = 1+q'r',\ \lambda_2 = 1-q'(1-r').$
As the integrand combined with the Jacobian contributes a factor of $q'^2$ this cancels the $(\lambda_1-\lambda_2)^{-2}=q'^{-2}$ contribution at $q'=0$ and the integrand is finite.
The integration domain is split in two parts: One disconnected rectangular part, where the integrations can be done independently from one another,
$q'\in[0,2],\ r' \in [0,1]$, and a connected part, where $q'\in[2,\infty),\ r' \in\left[(q'-2)/q',1\right]$. The characteristic function in the new variables reads $R_s(k)= R_s^{(d)}(k)+R_s^{(c)}(k)$, where
\begin{align}\nonumber
R_s^{(d)}(k)= 1 -\frac{k^2}{4}\int\limits_{0}^{2} dq'\int\limits_{0}^{1} dr' f(q',r')
   \end{align}
and 
\begin{align}\nonumber
 R_s^{(c)}(k)=-\frac{k^2}{4}\int\limits_{2}^{\infty} dq'\int\limits_{(q'-2)/q'}^{1}  dr' f(q',r') 
\end{align}
denote the disconnected and connected part of the characteristic function. The integrand is given by
\begin{align}\nonumber
f(q'&,r') \\\nonumber=\ &\mathcal{F}_U(1+q'r',1-q'(1-r'))\\\nonumber&\times
  \Biggl( \frac{r'(q'r'+2)}{(g_a^++ q'r'+1)(g_b^+ +q'r'+1)} \\\nonumber&+\frac{ (1-r')(2-q'(1-r'))}{(g_a^++1-q'(1-r'))}\\\nonumber&\times\frac{1}{(g_b^++1-q'(1-r'))}\Biggr)\\\label{eq:Rk}&\times  J_0\left( k \sqrt{\frac{q' r'(q'r'+2)}{(g_a^++q'r'+1)(g_b^+ +q'r'+1)}}\right).
\end{align} All odd moments vanish, which can be seen by expanding the Bessel function, see \cref{eq:Sub1}, in powers of $k$. All even moments necessarily exist due to the unitarity of the scattering matrix. The characteristic function also serves as moment generating function. We therefore have
\begin{align}
R_s(k) &=\label{eq:sum}\sum\limits_{n=0}^{\infty}(-1)^n\frac{k^{2n}}{(2n)!}\overline{x_s^{2n}}.
\end{align} The zeroth order moment is equal to one and corresponds to the Efetov-Wegner-contribution \cite{Efetov, Efetov1996, Kieburg2009}. In a related but different context, the channel factor was approximated in \cite{VerbaarschotCorr}  for infinitely many channels with $T_c\simeq 1/M$ in the form \begin{align}\mathcal{F}_U(1+q'r',1-q'(1-r')) \simeq \exp{(-\pi \Xi q' )}.\end{align} Here, we also use this limit to calculate the moments. The asymptotes for the moments can then be obtained by Watson's lemma \cite{Expansions, Laplace1, Laplace2, Wong} and integrating over $r'$. The asymptotically relevant part is the integration over $[0,2]\times [0,1]$ as this is the only part where $q'=0$ can be attained and the exponentially decaying channel factor becomes maximal. The moments associated with the characteristic function are asymptotically determined by 
\begin{align}\nonumber
\overline{x_s^{2n}}\\\nonumber
\simeq\ &\frac{\Gamma{(2n+1)}}{2^{2n}\Gamma^2(n)} \sum\limits_{m=0}^{\infty} \left(\frac{1}{\pi \Xi}\right)^{m+1}  \frac{\partial^m}{\partial q'^m} \int\limits_{0}^{1}dr'(q' r')^{n-1}\\\notag&\times \Biggl( \frac{r'(q'r'+2)}{(g_a^++ q'r'+1)(g_b^+ +q'r'+1)} \\\nonumber&+\frac{ (1-r')(2-q'(1-r'))}{(g_a^++1-q'(1-r'))(g_b^++1-q'(1-r'))}\Biggr)\\\label{eq:KorrekturMomente}&\times\left( \frac{q'r'+2}{(g_a^++q'r'+1)(g_b^+ +q'r'+1)} \right)^{n-1}  \Biggl|_{q'=0}.
\end{align}
After exchanging differentiation and integration, the derivatives over $q'$ are done via chain rule and the remaining integrand is a polynomial in $r'$.
Evaluating the leading order term of the moments yields 
\begin{align}\nonumber
\overline{x_s^{2n}} 
=\ & \frac{(2n)!}{n!} \left(\frac{1}{2(g_a^+ + 1)(g_b^+ + 1)\pi \Xi }\right)^n\\&+\mathcal{O}( \Xi^{-(n+1)}).
\end{align}
The explicit calculation is given in $\textit{Appendix A}$.
Inserting the first order asymptote into \cref{eq:sum} leads to
\begin{align}\nonumber
R_s(k) 
&\simeq \sum\limits_{n=0}^{\infty}(-1)^n k^{2n}\frac{1}{n!} \left(\frac{1}{2(g_a^+ + 1)(g_b^+ + 1)\pi \Xi }\right)^n\\\label{eq:erste}&= \exp{\left(-\frac{k^2}{2(g_a^+ + 1)(g_b^+ + 1)\pi \Xi}\right)}.
\end{align}
Thus, by taking the inverse Fourier transform of the characteristic function, we arrive at the Gaussian approximation of the distribution $P_s(x_s)$ for either the real part $x_1$ or imaginary part $x_2$ of $S$-matrix element.
We find
\begin{align}\nonumber
P_s(x_s)\simeq\ &\mathcal{F}^{-1}[R_s(k)](x_s) \\\nonumber=\ & \sqrt{\frac{ \Xi(g_a^++1)(g_b^++1)}{2}}\\\label{eq:Gaussian}&\times\exp{\Biggl(-\frac{\pi \Xi(g_a^++1)(g_b^++1)x_s^2}{2}\Biggr)}
\end{align}
for $s=1, 2$.
This is the Gaussian arising from the Ericson limit when neglecting higher order corrections to the moments. The contribution from the connected part of the integration domain is globally decaying with respect to $ \Xi$ and thus negligible. Put differently, a Gaussian arises in the Ericson regime if we consider a many channel limit, neglect the connected part  of the integration domain and the asymptotically negligible part at $q'=2$. 

\textit{Higher order corrections} ---
To second order in powers of $1/\Xi$, the 
 moments are found to be
\begin{align}\nonumber
\overline{x_s^{2n}}
=\ & \frac{(2n)!}{n!} \left(\frac{1}{2(g_a^+ + 1)(g_b^+ + 1)\pi \Xi }\right)^n\\\nonumber&+\frac{g_a^+g_b^+-g_a^+-g_b^+-3}{(2(g_a^++1)(g_b^++1)\pi \Xi)^{n+1}}\frac{\Gamma{(2n+1)}}{\Gamma{(n-1)}}\\&+\mathcal{O}( \Xi^{-(n+2)}).
\end{align} By resummation, the corresponding asymptote of the characteristic function reads
\begin{align}\nonumber
R_s(k)\simeq\ & \exp{\left(-\frac{k^2}{2(g_a^++1)(g_b^++1)\pi \Xi}\right)}\\\label{eq:FourierChar}&\times\left(1+\frac{(g_a^+g_b^+-g_a^+-g_b^+-3)k^4}{8((g_a^++1)(g_b^++1)\pi \Xi)^3}\right).
\end{align}
The associated correction to the distribution is thus
\begin{align}\nonumber
P_s(x_s)\simeq\ &\mathcal{F}^{-1}[R_s(k)](x_s)\\\nonumber=\ &\sqrt{\frac{ \Xi(g_a^+ + 1)(g_b^+ + 1)}{2}}\\\nonumber&\times\exp{\left(-\frac{\pi \Xi(g_a^+ + 1)(g_b^++1)x_s^2}{2}\right)}\\\nonumber&\times\Biggl(1+\Biggl(3-6\pi \Xi(g_a^+ + 1)(g_b^+ +1)x_s^2\\\nonumber&+(\pi \Xi(g_a^+ + 1)(g_b^+ + 1))^2x_s^4\Biggr) \\\label{eq:Zweite}&\times \frac{g_a^+g_b^+-g_a^+-g_b^+-3}{8(g_a^++1)(g_b^++1)\pi \Xi} \Biggr).
\end{align}
The asymptote of the distribution is proceeding in fractional powers of $1/\Xi$. For $\Xi\gg1$, the distribution in \cref{eq:Zweite} converges to a Gaussian, since the contributions around $x_s=0$ become negligible except for the term $(\Xi(g_a^++1)(g_b^++1)/2)^{1/2}$. If $x_s \neq 0$, each term in \cref{eq:Zweite} contains an exponential decay in $\Xi$ that ensures convergence, since 
$\exp{(-a\Xi)}\Xi^\mu,\ a,\mu>0,$ goes to zero in the large--$\Xi$ limit. Analogously to the Central Limit Theorem, a proper rescaling is called for to obtain pure $1/\Xi$  asymptotics for the corrections, see $\textit{Appendix B}$. The symmetry $P_s(x_s)=P_s(-x_s)$ is conserved by the asymptotic approximation as well as the symmetry $a\leftrightarrow b$ of the scattering channels. Around $x_s=0$, the correction to the Gaussian maximum is 
\begin{align}\nonumber
P_s(0)= \ &\sqrt{\frac{ \Xi(g_a^+ + 1)(g_b^+ + 1)}{2}}\\\nonumber&\times\Biggl(1+ \frac{3(g_a^+g_b^+-g_a^+-g_b^+-3)}{8(g_a^++1)(g_b^++1)\pi \Xi} \Biggr)\\\label{eq:CorrMax}&+\mathcal{O}( \Xi^{-3/2}).
\end{align}
The maximum of the distribution is thus not given by that of a Gaussian if $\Xi \approx 1$ and the resonances are only weakly overlapping. Depending on the scattering channels $T_a$ and $T_b$, the maximum of the distribution is shifted by the contribution $(3(g_a^+g_b^+-g_a^+-g_b^+-3))/(8(2(g_a^++1)(g_b^++1)\Xi)^{1/2}\pi)+\mathcal{O}(\Xi^{-3/2})$. This shift can be positive as well as negative. In view of \cref{eq:CorrMax}, a positive shift occurs if $1-T_a-T_b>0$ and a negative shift if  $1-T_a-T_b<0$. 

\begin{figure}[H]
	\centering
	\captionsetup[subfigure]{labelformat=empty}
\begin{subfigure}[b]{1\linewidth}
\centering
				\includegraphics[width=\linewidth]{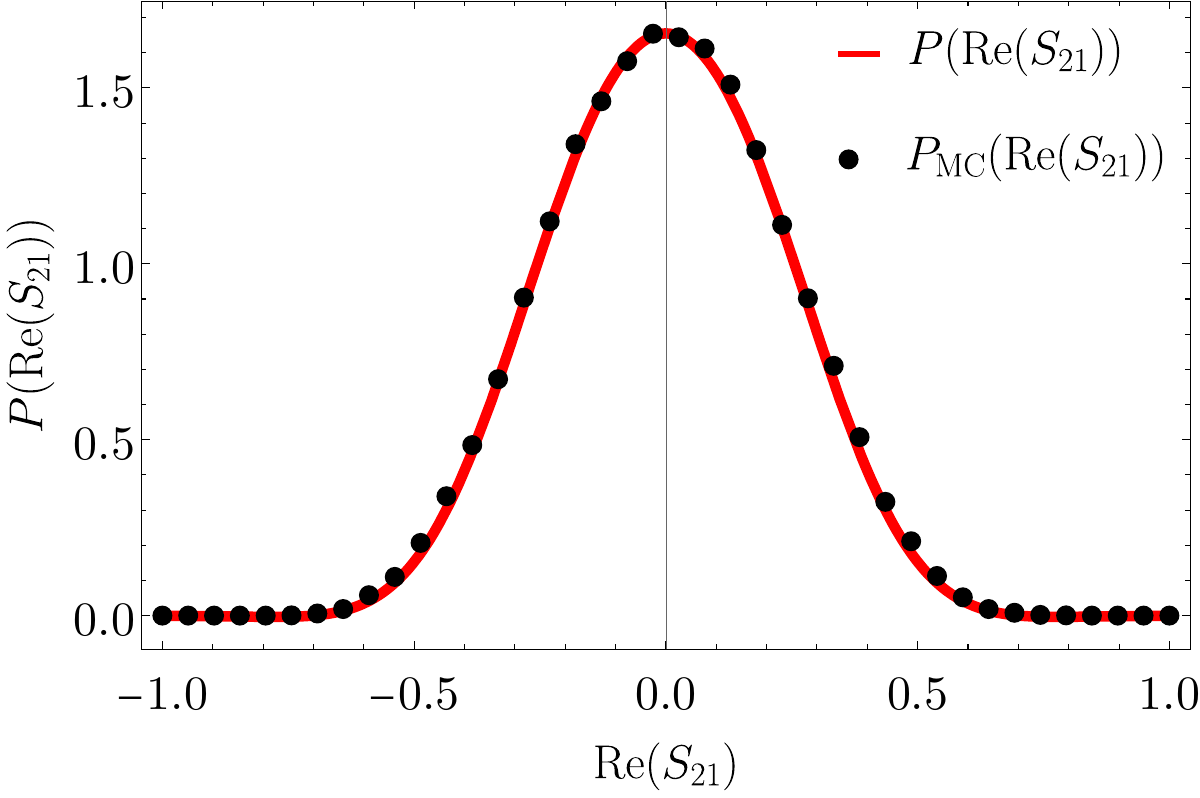} 
				\subcaption{}
\label[figure]{P1MC}
	\end{subfigure}
~
\begin{subfigure}[b]{1\linewidth}
\centering
	\includegraphics[width=\linewidth]{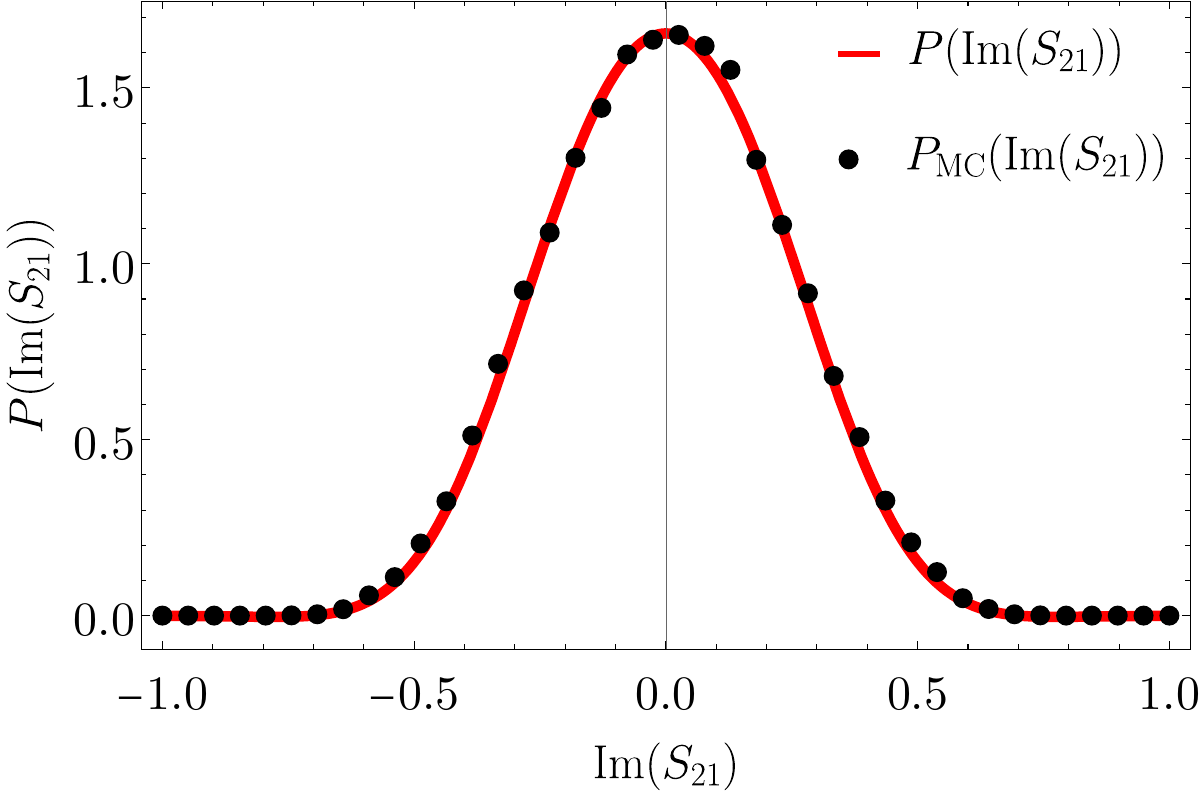} 
\subcaption{}
\label[figure]{P2MC}
	\end{subfigure}
~	
\begin{subfigure}[b]{1\linewidth}
\centering
	\includegraphics[width=1\linewidth]{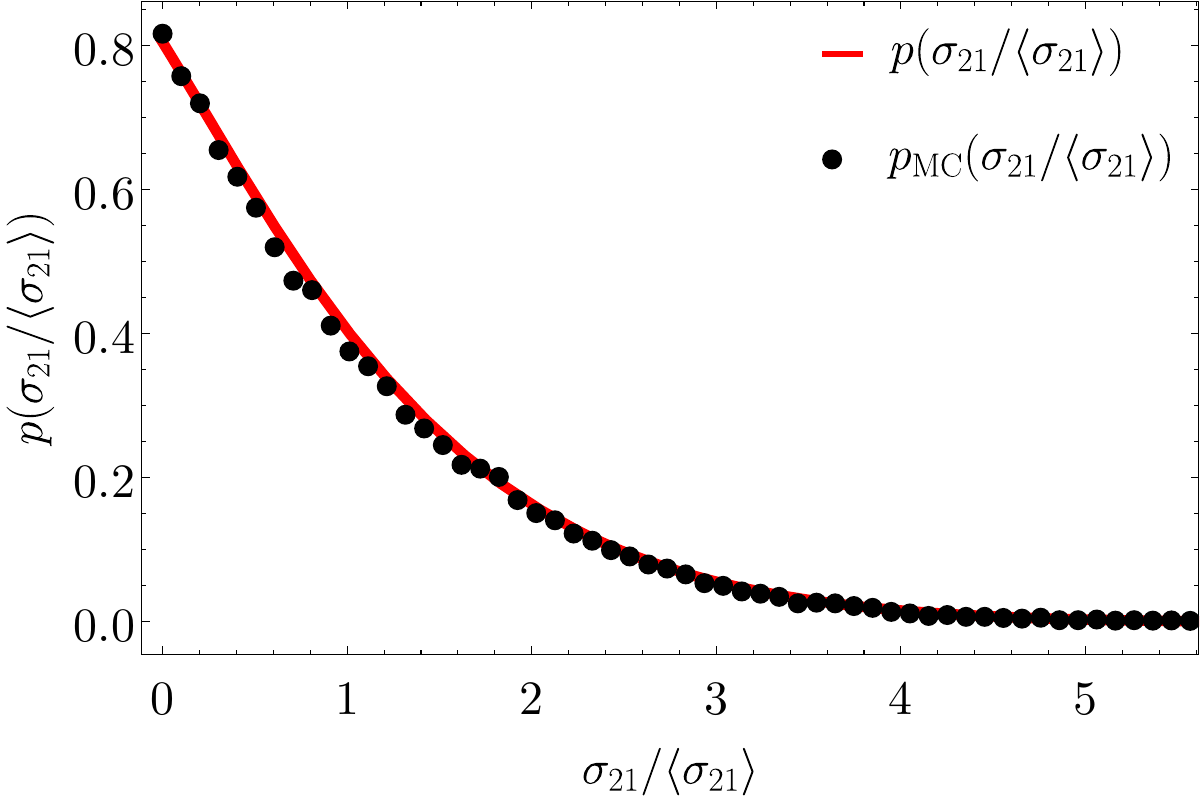} 
		\subcaption{}
\label[figure]{CrosssectionsMC}
\end{subfigure}	
\caption{Distribution of $\text{Re}(S_{21})$, $\text{Im}(S_{21})$ and normalized distribution of cross--section $\sigma_{21}$ for $\Xi=1.235$ and $M=52$.}
\label[figure]{DistributionsMC}
\end{figure}

\begin{figure}[H]
	\centering
	\captionsetup[subfigure]{labelformat=empty}
\begin{subfigure}[b]{1\linewidth}
\centering
				\includegraphics[width=\linewidth]{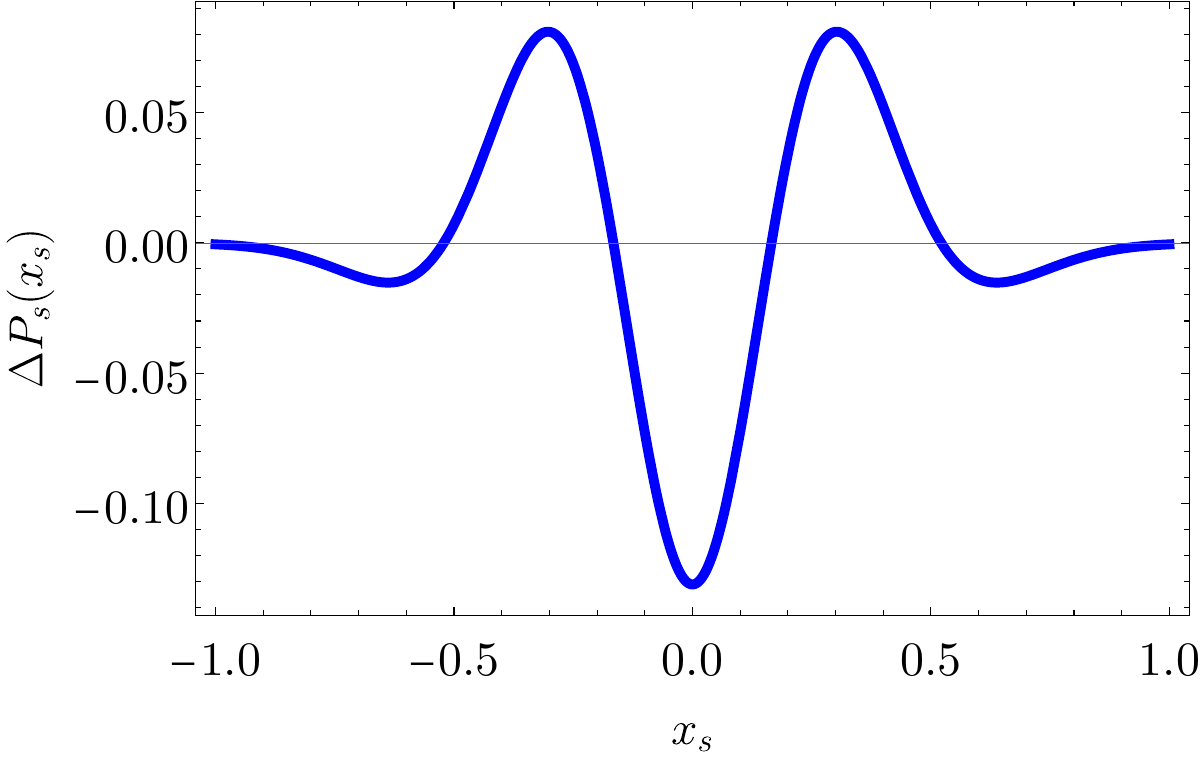} 
				\subcaption{}
\label[figure]{Difference1}
\end{subfigure}
~
\begin{subfigure}[b]{1\linewidth}
\centering
	\includegraphics[width=1\linewidth]{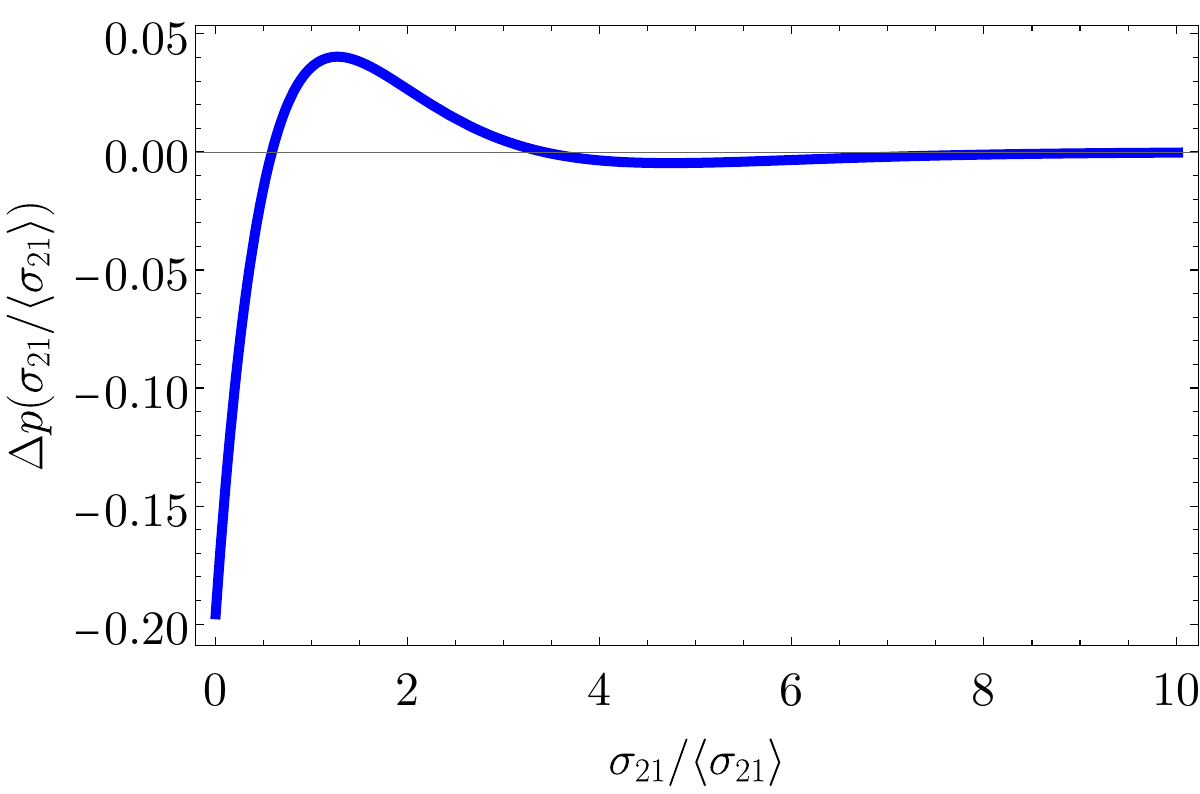} 
					\subcaption{}
\label[figure]{Difference2}
\end{subfigure}	
\caption{Differences $\Delta P_s(x_s)$ and $\Delta p(\sigma_{21}/\langle \sigma_{21} \rangle)$, where $x_1 = \text{Re}(S_{21}),\ x_2 = \text{Im}(S_{21})$, $\Xi=1.235$ and $M=52$.}
\label[figure]{Corrections}
\end{figure}

\begin{figure}[H]
	\centering
		\captionsetup[subfigure]{labelformat=empty}
	\begin{subfigure}[b]{1\linewidth}
				\includegraphics[width=1\linewidth]{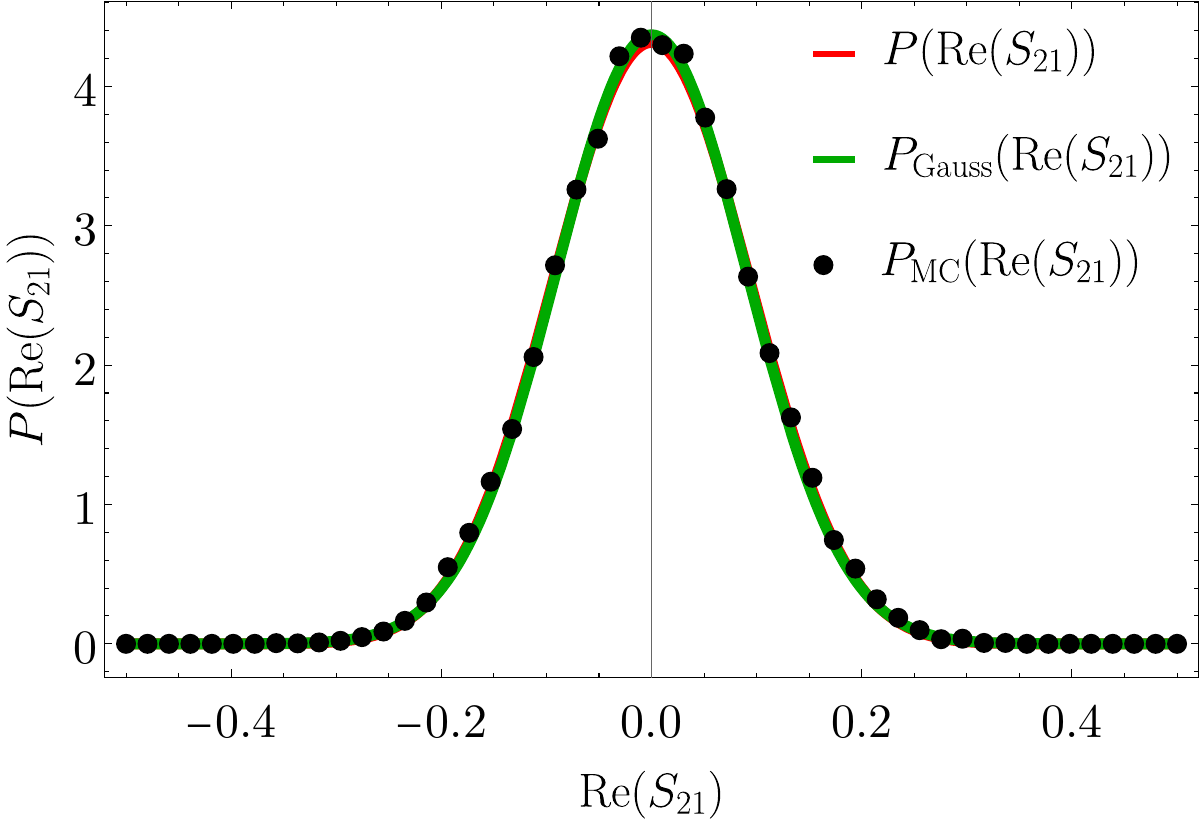} 
					\subcaption{}
\label[figure]{Gauss1}
\end{subfigure}
~
\begin{subfigure}[H]{1\linewidth}
	\includegraphics[width=1\linewidth]{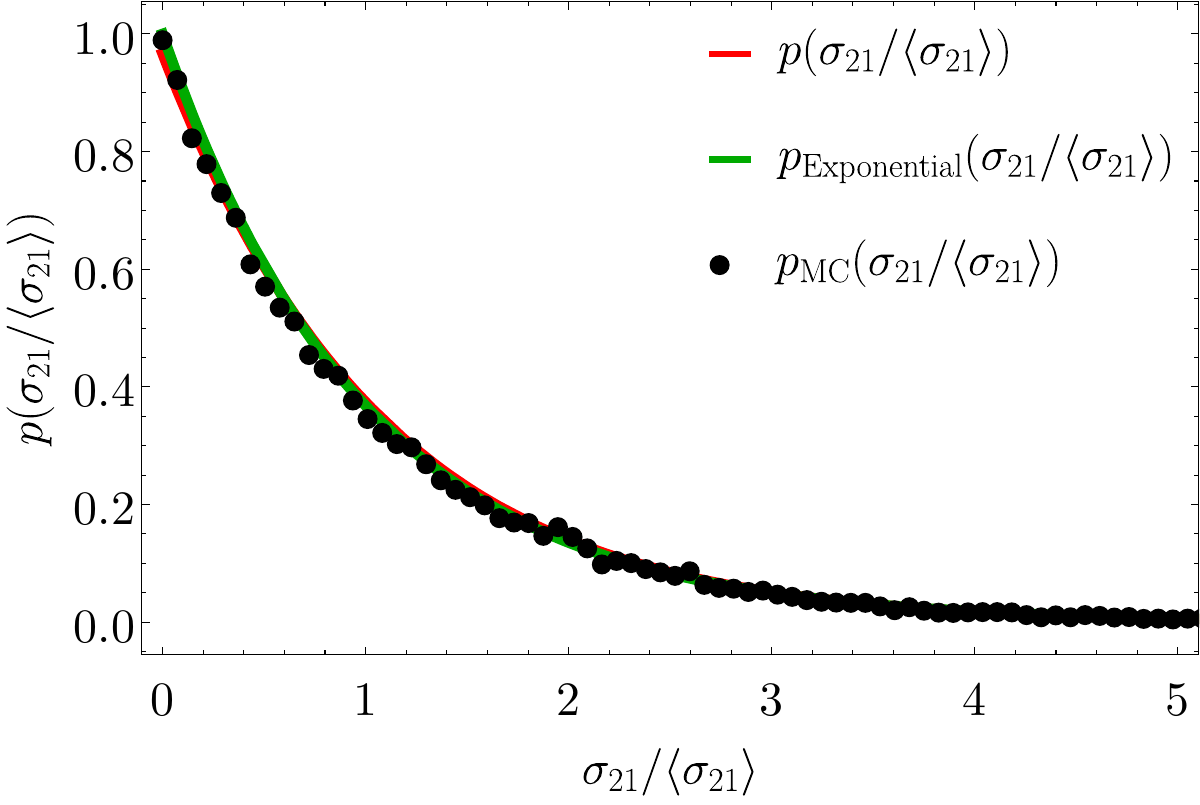} 
	\subcaption{}
\label[figure]{CS}
\end{subfigure}
\caption{Distribution of  $\text{Re}(S_{21})$ and normalized cross--section $\sigma_{21}/\langle \sigma_{21}\rangle $ for $\Xi=9.55$ and $M=60$.}
\label[figure]{Gauss}
\end{figure}

\textit{Relation to the cross--section distribution} ---
In many experiments only the cross--sections are accessible. 
The distribution of cross--sections $\sigma_{ab}= |S_{ab}|^2 = \text{Re}^2(S_{ab})+\text{Im}^2(S_{ab})$ is calculated from the bivariate characteristic function via a two-dimensional Fourier transform \cite{Wirkungsquerschnitte}.  By exploiting the radial symmetry of the bivariate characteristic function, the cross--section distribution can be written as the Hankel transform of the monovariate characteristic function, i.e.
\begin{align}
p(\sigma_{ab}) = \frac{1}{2}\mathcal{H}_0[R_s(k)](\sqrt{\sigma_{ab}}).
\end{align}
 Using the asymptotic expression in \cref{eq:FourierChar}, we arrive at
\begin{align}\nonumber
p(\sigma_{ab})=\nonumber
\frac{1}{2}&\exp{\left(-\frac{\sigma_{ab} (g_a^++1)(g_b^++1)\pi \Xi}{2}\right)}\\\nonumber\times&\Biggl((g_a^++1)(g_b^++1)\pi \Xi \\\nonumber+&(g_a^+g_b^+-g_a^+-g_b^+-3)\\\nonumber\times &\Biggl(1-(g_a^++1)(g_b^++1)\pi \Xi \sigma_{ab}  \\\label{eq:crosssections}+&\frac{1}{8}((g_a^++1)(g_b^++1)\pi \Xi \sigma_{ab})^2\Biggr)\Biggr).
\end{align}
As can be seen from \cref{eq:crosssections}, there is an additional contribution to the pure exponential decay expected in the Ericson regime. At $\sigma_{ab}=0$, we have
\begin{align}\nonumber
p(0)=\ &\frac{(g_a^++1)(g_b^++1)\pi \Xi}{2}\\\label{eq:shift}&+\frac{g_a^+g_b^+-g_a^+-g_b^+-3}{2} +\mathcal{O}( \Xi^{-1}).
\end{align}
The term $(g_a^+g_b^+-g_a^+-g_b^+-3)/2$ is the shift relative to the maximum of the exponential decay, which is attained at $\sigma_{ab}=0$. This quantitatively explains the observed deviations from the exponential decay in the onset of the Ericson regime \cite{Wirkungsquerschnitte}.
 \indent\textit{Comparison with experimental data and numerical simulations} ---
To further validate the asymptotic expressions, we compare our results with data from microwave measurements \cite{Quantengraphen}.  In the experiment under consideration, a microwave network is realized by using coaxial microwave cables coupled by joints. We have two strong scattering channels $a, b$ with $T_a=T_b=\num{0.88}$ and a parameter $\tau_{\text{abs}}=6$ to account for absorption \cite{Dietz2010}. To model the absorption, $50$ fictitious channels of equal transmission were used. Using the Weisskopf estimate
$
\Xi=(T_1+T_2+\tau_{\text{abs}})/(2\pi) \approx \num{1.235}
$
we find that the measurements are taken in the onset of the Ericson regime, but the resonances are weakly overlapping. We focus on the $S$-matrix element $S_{21}$.
\begin{figure}[H]
	\centering
	\captionsetup[subfigure]{labelformat=empty}
\begin{subfigure}[b]{1\linewidth}
\centering
				\includegraphics[width=\linewidth]{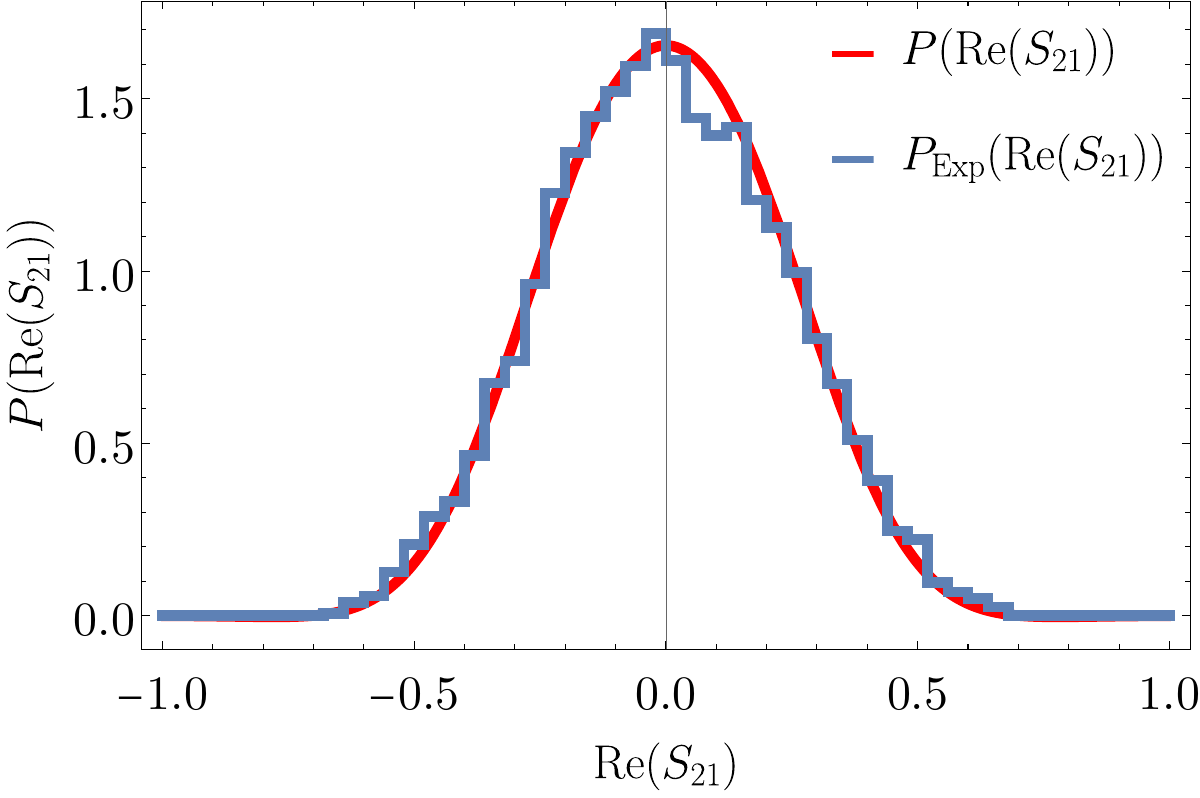} 
				\subcaption{}
\label[figure]{P1}
	\end{subfigure}
~
\begin{subfigure}[b]{1\linewidth}
\centering
	\includegraphics[width=\linewidth]{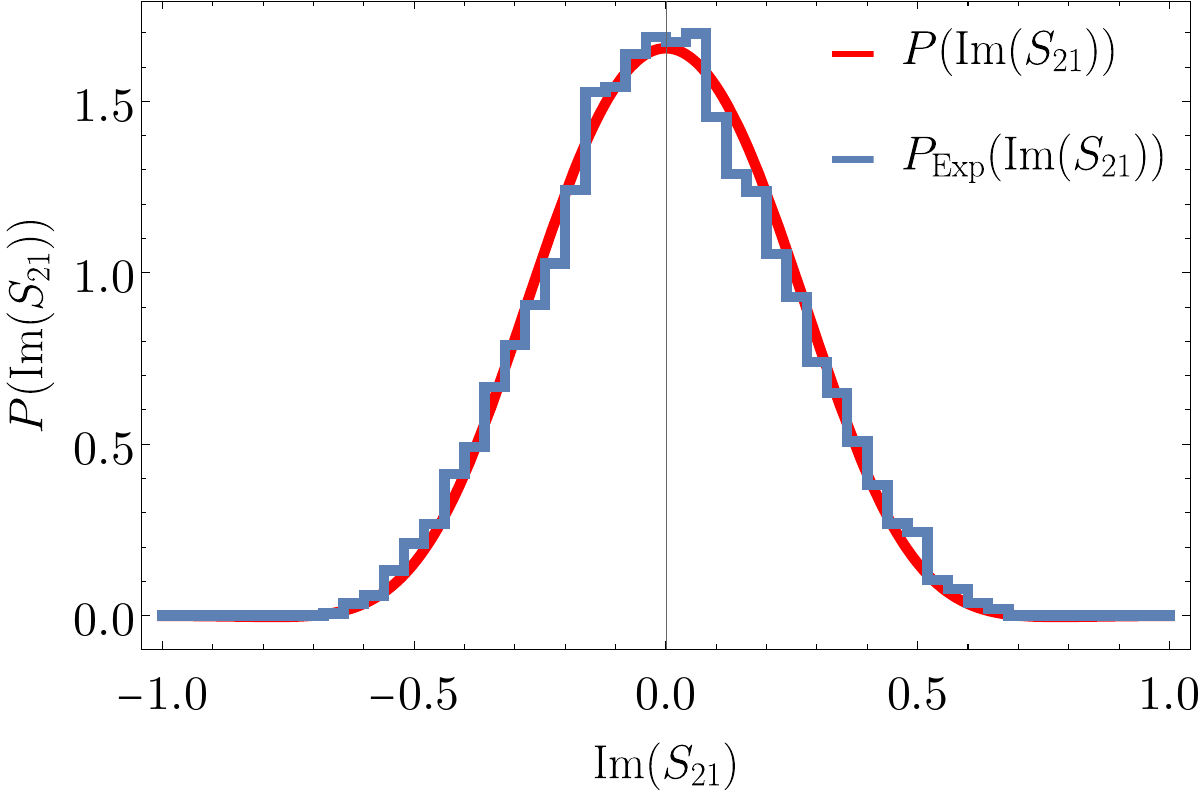} 
\subcaption{}
\label[figure]{P2}
	\end{subfigure}
~	
\begin{subfigure}[b]{1\linewidth}
\centering
	\includegraphics[width=1\linewidth]{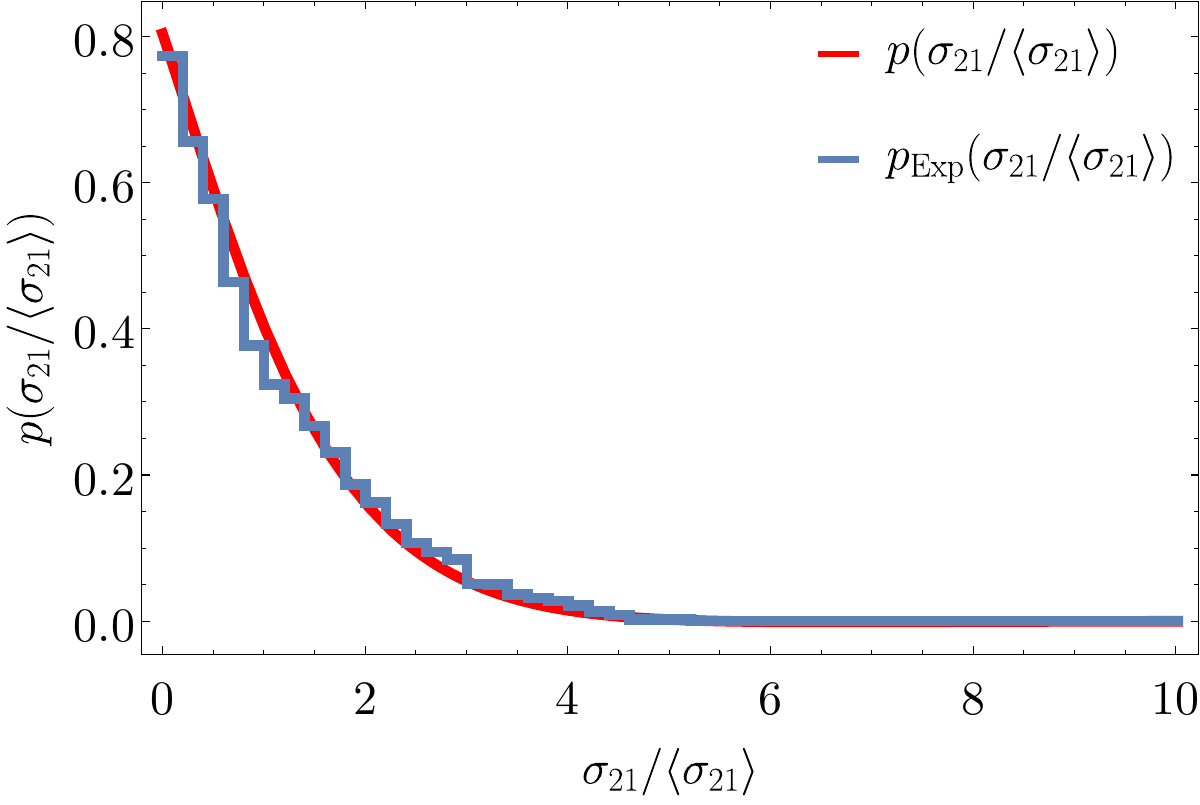} 
		\subcaption{}
\label[figure]{Crosssections1}
\end{subfigure}	
\caption{Distribution of $\text{Re}(S_{21})$, $\text{Im}(S_{21})$ and normalized distribution of cross--section $\sigma_{21}$ for $\Xi=1.235$ and $M=52$.}
\label[figure]{Distributions1}
\end{figure}
As seen in \cref{Distributions1}, the analytical result is in good agreement with the experimental data. The measured distribution is equal for real and imaginary part. The peak of the distribution is non-Gaussian. This behaviour is expected, since the data is obtained in the onset of the Ericson regime. For $\Xi \gg 1$, the distribution undergoes a transition to the Gaussian distribution given in \cref{eq:Gaussian}. The distribution of cross--sections deviates from the exponential, since the contribution at $\sigma_{ab}=0$ differs by the shift $(g_a^+g_b^+-g_a^+-g_b^+-3)/2 +\mathcal{O}(\Xi^{-1})$ given in \cref{eq:shift} compared to the pure exponential decay. 
 We also compare the analytical result with Monte Carlo simulations. The Monte Carlo simulations were done by sampling $30000$ random matrices with dimension $N=200$ from the GUE and numerically calculating the corresponding $S$-matrix elements by \cref{eq:Heidelberg}. As seen in \cref{DistributionsMC}, the analytical expressions compare well to the Monte Carlo simulations.
To visualize the effects before entering the Ericson regime the difference of the correction from \cref{eq:Zweite} and the Gaussian
\begin{align}\nonumber
\Delta P_s(x_s)\simeq\ &\sqrt{\frac{\Xi(g_a^+ + 1)(g_b^+ + 1)}{2}}\\\nonumber&\times\exp{\left(-\frac{\pi \Xi(g_a^+ + 1)(g_b^++1)x_s^2}{2}\right)}\\\nonumber&\times\Biggl(\Biggl(3-6\pi \Xi(g_a^+ + 1)(g_b^+ +1)x_s^2\\\nonumber&+(\pi \Xi(g_a^+ + 1)(g_b^+ + 1))^2x_s^4\Biggr) \\&\times \frac{g_a^+g_b^+-g_a^+-g_b^+-3}{8(g_a^++1)(g_b^++1)\pi \Xi} \Biggr)
\end{align}
is depicted in \cref{Corrections}.
The comparison for the correction to the cross--section \begin{align}\nonumber
\Delta p(\sigma_{ab})=\ &\frac{1}{2}\exp{\left(-\frac{\sigma_{ab} (g_a^++1)(g_b^++1)\pi \Xi}{2}\right)}\\\nonumber&\times(g_a^+g_b^+-g_a^+-g_b^+-3)\\\nonumber&\times \Biggl(1-(g_a^++1)(g_b^++1)\pi \Xi \sigma_{ab}  \\&+\frac{1}{8}((g_a^++1)(g_b^++1)\pi \Xi \sigma_{ab})^2\Biggr)\end{align} is also displayed in \cref{Corrections}. We normalize the cross--section distribution to its expectation value in the Ericson regime, $\langle \sigma_{ab} \rangle = 2/(\pi \Xi(g_a^++1)(g_b^++1))+\mathcal{O}(\Xi^{-2})$.
The sizeable dip at zero in \cref{Corrections} manifests for weakly overlapping resonances the significant deviations from the Gaussian. The distribution is also broadened in the tails. 
 The cross--section distribution on the other hand shows a maximal deviation from the exponential decay at zero, which is clearly visible in \cref{Corrections}. 
Finally, we  test our results for a set of parameters which is deeply in the Ericson regime numerically. We choose $T_c=1$ for $M=60$ channels, which corresponds to $\Xi\approx 9.55$. By doing so, the corrections in $1/\Xi$ become small compared to the leading order term and we indeed find Gaussian and exponential behavior, respectively, as seen in \cref{Gauss}. 
\indent\textit{Conclusion} ---
We succeeded in quantitatively capturing the Ericson transition of non-time-reversal invariant systems within the Heidelberg approach. We gave a full description how the Gaussian limit arises from the microscopic model and managed to calculate asymptotic series for the probability distribution of the off-diagonal $S$-matrix elements and cross--sections. We found excellent agreement of the analytical results with experimental data and numerical simulations in the transition region from weakly to strongly overlapping resonances. The methods developed here also carry over to the case of orthogonally and symplectically
invariant Hamiltonians. As the calculations become technically much more entangled, this material is beyond the scope of this contribution and will be presented elsewhere. This will allow for characterization of the Ericson transition also in time-reversal invariant cases. 
\\ \indent\textit{Acknowledgments} ---
We are grateful to Nils Gluth for fruitful discussions. Two of us (JC, BD) acknowledge financial support for the experiments from the China National Science Foundation (NSF), grants 11775100, 12247101 and 11961131009. One of us (BD) is grateful for funding from the Institute for Basic Science in Korea, project IBS–R024–D1.
\twocolumngrid
\bibliography{EricsonTransition}
\appendix
\onecolumngrid
\section{Endmatter}
\textit{Appendix A: Explicit derivation of the  Gaussian} ---
  Here, we explicitly calculate the integrals and derivatives arising in the leading order term 
 of the 2nth-moment. Splitting the integral from \cref{eq:KorrekturMomente} into two parts and only calculating the leading order term ($m=n-1$), we have
  \begin{align}\nonumber
\int\limits_{0}^{1}dr'\frac{\partial^{n-1}}{\partial q'^{n-1}} q'^{n-1}\left(\frac{r'(q'r'+2)}{(g_a^+ + q'r'+1)(g_b^+ + q'r'+1)}\right)^n \Biggl|_{q'=0} &= \int\limits_{0}^{1}dr'2^n (n-1)!\left( \frac{r'}{(g_a^+ +1)(g_b^+ +1)}\right)^n \\&= 2^n (n-1)!\left( \frac{1}{(g_a^+ +1)(g_b^+ +1)}\right)^n \frac{1}{n+1}.
\end{align}
The second integral reads
\begin{align}\nonumber
&\int\limits_{0}^{1}dr'\frac{\partial^{n-1}}{\partial q'^{n-1}} q'^{n-1}\Biggr[ \left( \frac{r'(q'r'+2)}{(g_a^+ + q'r'+1)(g_b^+ + q'r'+1)}        \right)^{n-1} \\\nonumber&\ \ \ \times\left(\frac{(1-r')(2-q'(1-r'))}{(g_a^+ +1-q'(1-r'))(g_b^++1-q'(1-r'))}\right)\Biggr] \Biggl|_{q'=0}\\\notag&= \int\limits_{0}^{1}dr' (n-1)! \left(\frac{2r'}{(g_a^+ + 1)(g_b^+ + 1)}\right)^{n-1}\frac{2(1-r')}{(g_a^+ + 1)(g_b^++1)}\\
&= 2^n (n-1)! \left(\frac{1}{(g_a^+ + 1)(g_b^+ + 1)}\right)^n \left(   \frac{1}{n} -\frac{1}{n+1}     \right)=  2^n (n-1)! \left(\frac{1}{(g_a^+ + 1)(g_b^+ + 1)}\right)^n\frac{1}{n(n+1)}.
\end{align}
All in all, the leading order asymptotic expression for the moments is
\begin{align}\nonumber
\overline{x_s^{2n}} =\ & \frac{\Gamma{(2n+1)}}{2^{2n}\Gamma^2(n)} \left(\frac{2}{2\pi \Xi}\right)^{n}\Biggl( 2^n (n-1)!\left( \frac{1}{(g_a^+ +1)(g_b^+ +1)}\right)^n \frac{1}{n+1}\\\notag& +2^n (n-1)! \left(\frac{1}{(g_a^+ + 1)(g_b^+ + 1)}\right)^n\frac{1}{n(n+1)}\Biggr)\\=\ &  \frac{\Gamma{(2n+1)}}{\Gamma(n)n}\left(\frac{1}{2(g_a^+ + 1)(g_b^+ + 1)\pi \Xi }\right)^n= \frac{(2n)!}{n!} \left(\frac{1}{2(g_a^+ + 1)(g_b^+ + 1)\pi \Xi }\right)^n.
\end{align}
By using the moment generating property of the characteristic function, we arrive at
\begin{align}\nonumber
R_s(k) &= \sum\limits_{n=0}^{\infty}(-1)^n k^{2n} \frac{1}{(2n)!} \frac{(2n)!}{n!} \left(\frac{1}{2(g_a^+ + 1)(g_b^+ + 1)\pi \Xi }\right)^n\\\label{eq:erste}
&= \sum\limits_{n=0}^{\infty}(-1)^n k^{2n}\frac{1}{n!} \left(\frac{1}{2(g_a^+ + 1)(g_b^+ + 1)\pi \Xi }\right)^n = \exp{\left(-\frac{k^2}{2(g_a^+ + 1)(g_b^+ + 1)\pi \Xi}\right)}.
\end{align}
Performing the inverse Fourier transform yields \cref{eq:Gaussian}. Higher order corrections are similarly calculated.

\textit{Appendix B: Rescaled asymptotics of $P_s(x_s)$} ---
By rescaling $x_s$ according to $\xi = \sqrt{\Xi}x_s$ and using \cref{eq:Zweite}, we find
\begin{align}\nonumber
	P_s(\xi)\simeq&\ \frac{1}{\sqrt{\Xi}}\mathcal{F}^{-1}[R_s(k)]\left(\frac{\xi}{\sqrt{\Xi}}\right)\\\nonumber=&\ \sqrt{\frac{ (g_a^+ + 1)(g_b^+ + 1)}{2}}\exp{\left(-\frac{\pi (g_a^+ + 1)(g_b^++1)\xi^2}{2}\right)}\\\nonumber&\times\Biggl(1+\Biggl(3-6\pi (g_a^+ + 1)(g_b^+ +1)\xi^2+(\pi (g_a^+ + 1)(g_b^+ + 1))^2\xi^4\Biggr) \\&\times \frac{g_a^+g_b^+-g_a^+-g_b^+-3}{8(g_a^++1)(g_b^++1)\pi \Xi} \Biggr).
\end{align}
At $\xi=0$, we have
\begin{align}
	P_s(0)= \sqrt{\frac{ (g_a^+ + 1)(g_b^+ + 1)}{2}}\left(1+  \frac{3(g_a^+g_b^+-g_a^+-g_b^+-3)}{8(g_a^++1)(g_b^++1)\pi \Xi}\right) +\mathcal{O}( \Xi^{-2}).
\end{align}
We see that the first correction to the Gaussian maximum is of order $1/\Xi$. The correction becomes negligible in the limit of large $\Xi$, which yields Gaussian behavior of the distribution. 
\end{document}